\documentclass[showpacs]{revtex4}

\usepackage{graphics}
\usepackage{amssymb}
\usepackage{amsmath}
\usepackage{soul}
\usepackage{color}
\usepackage{graphics}
\usepackage{amsmath,amssymb,amsfonts}
\usepackage{amssymb} 
\usepackage{bm}
\usepackage{latexsym}
\usepackage{ulem}

\begin{document}

\title{Physical Significance of Noether Symmetries }

\author{ Asghar Qadir }

\affiliation{
Abdus Salam School of Mathematical Sciences (ASSMS) Government College University, 68B New Muslim Town, Lahore, Pakistan} %
\email{asgharqadir46@gmail.com}

\author{Ugur Camci }

\affiliation{ Department of Chemistry and Physics, Roger Williams University, One Old Ferry Road, Bristol,  Rhode Island \mbox{02809, USA} }  
\email{ucamci@rwu.edu, ugurcamci@gmail.com}

\bigskip

\date{\today}


\begin{abstract} In this paper we will trace the development of the use of symmetry
in discussing the theory of motion initiated by Emmy Noether in 1918. Though
it started with its use in Classical Mechanics, and has been heavily used in
engineering applications of Mechanics, it came into its own in Relativity,
and Quantum Theory and their applications in Particle Physics and Field
Theory. It will be beyond the scope of this article to explain the Quantum
Field Theory applications in any detail, but the base for understanding it
will be provided here. We will also go on to discuss an insight from some
more mathematical developments related to Noether symmetry.
\end{abstract}



\maketitle

\section{Introduction}
\label{INT}

\noindent Physics started as a study of motion in Greek times and was
formalized by Aristotle (c. 350 BC) by a set of ``laws", which he declared as
``self-evident truths", based on his view of the Universe as it was then
visualized. The law for motion on the Earth was based on the nature of the
object moving (how much of Earth, Water, Air or Fire there is in it), and for
motion in the heavens on the ``truth" that heavenly objects are made of the
perfect element, aether, and hence move along perfect circles, unless they
are contaminated by proximity to the Earth, in which case epicycles (perfect
circles about a point moving in a perfect circle, or further repetitions
thereof) develop. The closer the object to the Earth, the more epicycles it
will have. Using this law the motion of the then known ``planets": the Sun,
Moon, Venus, Mars, Mercury, Jupiter and Saturn, were supposedly explained
using 127 epicycles. The number was reduced to 57 by Ptolemy (c. 150 AD) by
extending from motion in a plane to motion in three-dimensional space and
using spheres in place of circles. The Muslim scholars followed the same way
of thinking, but Al-Zarkali (c. 1050 AD) was ready to use ellipses instead of
perfect circles. Later this work, and the astronomical data of Ulugh Beg (c.
1420 AD), led Nicholas Copernicus to heliocentric planetary orbits \cite{nc}
in 1543 AD, with the Earth replacing the Sun as a planet. Johannes Kepler
\cite{jk} replaced Copernicus' circular orbits by ellipses in 1619, which
finally led Isaac Newton to his laws of motion and universal gravitation
\cite{in} in 1687. While Newton's law is supposedly universal, his methods
work well only for two bodies and become unwieldy for several bodies. The
method was extended by Joseph Louis Lagrange in 1778 and 1779, \cite{jll1}
and later used by William Rowan Hamilton \cite{wrh,wrh2} in 1834 and 1835, for
systems of particles. Newton had used Calculus for his purpose and Lagrange
used the Calculus of Variations that Leonhard Euler \cite{le} had fully
developed by 1773. It was this formulation that Emmy Noether had used, which
completed the classical view and led to the modern view of Mechanics.

Emmy Noether (1882-1935) was a German female - and in those days German
females could not enter academics. However, she had remarkable mathematical
capability and was able to get the support of David Hilbert to work in the
field unofficially. Finally she had to emigrate to America to achieve her
true potential. She had numerous contributions in various branches of
Mathematics, but we are concerned with her contribution to the use of
symmetry in Mechanics.

\newpage

In normal parlance ``symmetry" is used in an aesthetic sense to express
balance and harmony, as in William Blake's poem, {\it The Tyger}:\\
``Tyger, tyger burning bright;\\
In the forests of the night;\\
What immortal hand or eye\\
Could frame thy fearful symmetry?"\\
Almost as common is its use in the geometric sense of leaving a figure or
shape invariant under some transformation, such as reflection or rotation. It
is odd, however, that the common use has such a strong hold that many do not
realise that one can count the symmetries of objects. (AQ has even known an
expert in Differential Topology to objet to the idea of counting symmetries.)

The geometrical concept of symmetry had come from the Greeks. However, while
considering solutions of polynomial equations of degree 5 or more in 1771,
Lagrange extended the concept to invariance of polynomials under permutations
of its elements \cite{jll2}. While this led to many other developments in
Algebra, I am here concerned with its use by Abel \cite{nha} and Galois
\cite{eg} to invent Group Theory, so as to prove that there could be no
solution of quintic or higher degree polynomial equations by means of
radicals. In particular, Abel's work led to the Abelian Group and Galois' to
the Galois Group. This inspired Sophus Lie (1842-1899) to try to emulate the
success of Abel and Galois for differential equations in 1883 \cite{sl1}.
Note the leap over all other algebraic equations to reach out to {\it all}
differential equations. This was overambitious and Lie never managed to
complete the attempt. Nevertheless, it led to enormous developments in the
solution of nonlinear differential equations. This will be discussed in the
next section.

One might think that Geometry and Dynamics had no contact till the time that
Albert Einstein and Marcel Grossmann used Geometry to generalize the
Restricted Theory of Relativity \cite{aq1}, but that is not the case at all.
As pointed out by Julian Barbour \cite{jbb}, starting in antiquity and going
through Copernicus, Kepler, Galileo and Newton, Kinematics and then Dynamics,
have been inextricably entwined with Geometry. At the base of the link
between them is the idea of symmetry. Aristotle insisted on perfect circles
because he perceived them as the most symmetric figures possible. Ellipses,
to the contrary, were perceived as imperfect, and hence not to be used for
celestial motion. It took a deeper, hidden, symmetry for the ellipses to be
perceived as ``beautiful" by Kepler. We will see how the hidden symmetry was
uncovered by Noether. This symmetry could not have been understood till the
advent of calculus and it was the Geometry that used Calculus, Differential
Geometry, that Einstein and Grossmann introduced into considerations of
Dynamics. We will be concerned with the importance of the not-so-obvious
symmetries that have become all important in Modern Physics.

The plan of the paper is as follows. In the next section we will briefly review Lie's {\it Symmetry Analysis} and go on, in the subsequent sections three and four, to review Euler's variational principle for particles and fields, respectively, and its use by Lagrange and Hamilton in the principle of least action and the equations of motion of Hamilton and of Euler and Lagrange.  A geometrical
application is given in section five, to generalize the concept of a straight
line in flat space to curved spaces. The applications of the Noether's theorem in Classical Mechanics, Economics, Classical Field Theory, Relativistic Field Theory, will be discussed in the section six. Some extensions in obtaining Noether symmetries and Noether invariants will also be given in this section. Some applications of Quantum Field theory will be given in the next section.
Complex Lie and Noether symmetries will be considered in the section eight, and a discussion and conclusion presented in the last section.


\section{Lie Symmetry Analysis}
\label{lie-sym}

\noindent Before Lie, the usual method to solve a differential equation (DE) was by ad-hoc approaches or by approximating it by a linear DE and solving that. In general the approximation will work well enough in some domain and become arbitrarily bad in others. Thus one would need to prove the existence of a solution and to determine the domain in which the approximation is good enough. Since these will
be different for each DE, one is reduced to solving one DE at a time and cannot rely on any method for whole classes of DEs. Among the methods that had been used for solving nonlinear DEs was the transformation of independent and dependent variables. Analogous to Abel and Galois, Lie looked for invariance of the DE under such transformations \cite{sl1,sl2,sl3,sl4}, so that it could be determined when the DEs could be solved, or their order reduced, by transformation, and then proceed with the transformation. Lie used not only the groups of symmetries, but the algebra of the corresponding infinitesimal symmetry generators. The DEs are not necessarily single (scalar) but could be systems of (vector) DEs. Further, he did not restrict the domain of the DEs to be real, but took it to be complex.

For completeness we start with basic definitions so as to present the notation used. If there are $l$ independent variables represented as a vector ${\bf x}$ and $m$ dependent variables represented by ${\bf y}$, a {\it Lie point symmetry generator} is the operator
\begin{equation}
  {\bf X} = {\bf A} ({\bf x},{\bf y}) \cdot {\bf \nabla}_{{\bf x}} +  {\bf B} ({\bf x},{\bf y}) \cdot {\bf \nabla}_{{\bf y}} \, , \label{eq1}
\end{equation}
or using indices $a$ for the independent variables and $i$ for the dependent variables
\begin{equation}
  {\bf X} = A^a ( x^b, y^i) \frac{\partial}{\partial x^a} + B^i ( x^b, y^j) \frac{\partial}{\partial y^i} \, , \, ( a, b,... =1,...,l; i,j,...=1,...m) \, , \label{eq2}
\end{equation}
where the Einstein summation convention, that repeated indices are summed over, has been used. Further, if the DE is of order $n$, one must prolong the space and the generators to incorporate all the derivatives of the dependent variables with respect to the independent variables. For ordinary differential equations (ODEs)
\begin{equation}
  {\bf X}^{[n]} = A ( x, y^j) \frac{\partial}{\partial x} + B^i ( x, y^j) \frac{\partial}{\partial y^i} + B^{i\,[1]} ( x, y^j, {y^j}') \frac{\partial}{\partial y^i} + ... \, , \,\, ( a =1,...,l; i=1,...m) \, , \label{eq3}
\end{equation}
where
\begin{equation}
  B^{i\,[k]} = D_x B^{i\,[k-1]} - {y^i}' D_x A \, , \quad (k=1,...,m) \, , \label{eq4}
\end{equation}
$B^{i\,[0]}$ simply being $B^i$, and $D_x$ is the {\it total derivative in the prolonged space},
\begin{equation}
  D_x = \frac{\partial}{\partial x} + {y^i}' \frac{\partial}{\partial y^i} + ... + y^{i(k)} \frac{\partial}{\partial y^{i(k-1)}} \, . \label{eq5}
\end{equation}
For partial differential equations (PDEs), $A$ has to be replaced by ${\bf A}$ and $\partial / \partial x$ by $\nabla_{\bf x}$ in Eq. \eqref{eq3}. While the former is easily converted to index notation as $A^a$ and $\partial / \partial x^a$ in Eq. \eqref{eq3}, for the latter one has to write $y^{i\,[k]}$, which is the partial derivative with respect to {\it all} $x^a$ to {\it all orders} up to $k$.

The set of all prolonged symmetry generators, ${\bf X}_q \, (q = 1, ..., p)$, forms a $p$-dimensional
Lie algebra, which determines what reduction there can be of the DE. It is a Lie algebra if the commutators of the symmetry generators satisfy $[ {\bf X}_r, {\bf X}_s] = C^q_{r s} {\bf X}_q$ where $ C^q_{r s}$ are constants, called {\it structure constants}, that determine the structure
of the Lie algebra. The generators must also satisfy the Jacobi identities,
\begin{equation}
[[{\bf X}_{[i},X_j],X_{k]}] = 0~. \label{jacobi}
\end{equation}
The square brackets in the subscript denote a skew linear combination, signifying that the terms with the suffices in an even permutation are positive and those in an odd permutation negative, so that interchanging any two indices reverses the sign of the expression. The total expression is
divided by the factorial of the number of indices involved.
A system of $m$ ODEs of order $n$,
$$E^i (x, y^j, {y^j}',...,y^{j\,[n]} ) = 0,$$
is said to be {\it symmetric} under the transformation generated by $\bf{X}$, if ${\bf X}^{[n]} E^i = 0$, when restricted to the solutions of $E^i = 0$, which is denoted by ${\bf X}^{[n]} E^i = 0|_{{\bf E}=0}$. The generalization to PDEs is as before, with the corresponding complications.

It is worth pondering the prolonged space. For a scalar ODE there are only two variables, the independent and dependent, and so the manifold considered is only two-dimensional. However, for a DE we have to treat the derivative as unknown, for if it were known the DE would already be reduced or solved. Hence it must be treated as another dependent variable, leading to a three-dimensional manifold. If the DE is of second order, we would need to also treat the second derivative as
an independent variable. Thus, for an $n^{\rm th}$ order ODE, we need to use an $(n + 2)-$
dimensional space. This is the {\it prolonged} or {\it extended space}, also called a {\it jet space}. For an $m$-dimensional system of $n^{\rm th}$ order ODEs we need an $(n m + m + 1)-$dimensional space. For a PDE of $l$ independent variables the dimension is $(n m l + m + l)$. Since the largest group acting on an $n-$dimensional manifold is $GL(n)$, that would be the upper limit for the algebra. In fact, since the algebra of infinitesimal generators must leave the group identity out, and only add to it, the relevant group would be $SL(n)$ and the corresponding algebra $sl(n)$, which has $n(n - 1)/2$ generators. Thus the dimension for the general PDE mentioned is $(m n l + m + l)(m n l + m + l - 1)/2$. Obviously, the dimension of the manifold rapidly becomes unwieldy with increase in order, and the number of independent and dependent variables. Even for a second order, two dimensional PDE of two variables the dimension is $66$, and this ignores other complications arising for PDEs, which will be discussed later. This is what makes it necessary to use algebraic computing programmes.

\section{The Variational Principle for Particles}
\label{variation}

\noindent The laws of refraction had been explained by assuming that light takes minimum time to go from a point in one medium to a point in the other medium, taking into account the change in its speed in the media. This principle was used by Lagrange \cite{jll1} for his extension of Newton's mechanics. He wrote Newton's mechanics for a system of particles as if it were a single particle in a higher dimensional space. For $N$ particles, Newton's laws would be written in terms of their positions, ${\bf r}_i (t)$, and velocities ${\bf \dot{r}}_i (t)$ ($i=1,...,N$). Lagrange wrote them as $ (q^a (t), \dot{q}^a (t) ), (a =1,...,3 N)$. This change makes it possible to also incorporate m constraints between the generalized coordinates, so that the dimension of the space for the single particle is $n = 3N - m$. He then required that the {\it free energy}, i.e. the difference between the kinetic and potential energy, be minimum, over the entire motion. This is called the {\it principle of least action}.

Lagrange's methods have been used in areas far removed from Mechanics, like Economics. The function to be minimized (with or without constraints) is called the {\it Lagrangian}, $L [t, q^i (t), \dot{q}
^i (t)]$, and the {\it action} is defined as the total of the minimized quantity over the time period from the initial time, $t_i$, to the final time, $t_f$ :
\begin{equation}
  S \left[ q^i (t), \dot{q}^i (t) \right] = \int_{t_i}^{t_f} \mathcal{L} \left[ t, q^i (t), \dot{q}^i (t)  \right] dt \qquad (i = 1,...,n).
\end{equation}
In Economics, the $q^i (t)$ would represent the quantity of a commodity and its time rate of change would be directly related to the price of that commodity in the market. The Lagrangian would be the cost for all things bought at the time, and the action would be the total money spent. It is useful to think of this analogy for Mechanics. In that case the money spent at one instant is the free-energy and the total money spent is the total energy spent. The object ``wants" to spend the least energy to get from the start to the end, and ``chooses" the path that will do this.

The original mechanical application of Lagrange's formalism regarded the Lagrangian as the difference between the kinetic and potential energy
\begin{equation}
  \mathcal{L} \left[ t, q^i (t), \dot{q}^i (t)  \right] = T ( \dot{q}^i (t) ) - V( q^i (t) )  \, ,  \label{lagr-1}
\end{equation}
where the explicit time-dependence is absent. This was because in the system of the Sun and planets the gravitational potential remained constant, and there was no meaning to the kinetic energy changing with time. The Economic analogy replaces the kinetic energy by the profit and the potential energy by the loss. In Economics, we are familiar with money ``evaporating" due to inflation. Similarly, the potential can be made time-dependent and energy can be lost to friction, or be radiated away. Hence there should be explicit time dependence. Not only may the system lose energy, it could gain energy. The corresponding phenomenon in Economics is ``deflation", where the money appreciates in value. One might think that would be desirable, but it leads to the Economy slowing down. In Physics, one gets more energy, but it is unusable as it is thermalized; it would lead to a ``heat death".

For the action functional, $S$, to be minimal when we vary the functions that are its
arguments by $\delta q^i, \delta \dot{q}^i$, it must be unchanged, i.e. $\delta S = 0$. Writing this explicitly,
\begin{equation}
  \delta S = \int_{t_i}^{t_f} \left[ \frac{\partial \mathcal{L}}{\partial q^i} \delta q^i + \frac{\partial \mathcal{L}}{\partial \dot{q}^i} \delta \dot{q}^i  +  \frac{\partial \mathcal{L}}{\partial t} \delta t  \right] dt = 0 \, .
\end{equation}
If we require that the initial and final positions of the system of particles are fixed,
then $\delta q^i (t_i) = \delta q^i (t_f) = 0$. Writing the variations in the integrand in terms of $\delta q^i$ and integrating the $\delta \dot{q}^i$ by parts using the above boundary conditions,
\begin{equation}
  0 = \int_{t_i}^{t_f} \left[ \frac{\partial \mathcal{L}}{\partial q^i} - \frac{d}{d t} \left( \frac{\partial \mathcal{L}}{\partial \dot{q}^i} \right) + \frac{1}{\dot{q}^i } \frac{\partial \mathcal{L}}{\partial t} \right] \delta q^i dt \, .
\end{equation}
Since the $\delta q^i$ in the integrand is arbitrary and the integral is zero the rest of the
integrand must be zero. This gives the time-dependent Euler-Lagrange (EL) equations for a system of particles
\begin{equation}
  \frac{d}{d t}  \left( \frac{\partial \mathcal{L}}{\partial \dot{q}^i} \right) = \frac{\partial \mathcal{L}}{\partial q^i} + \frac{1}{\dot{q}^i } \frac{\partial \mathcal{L}}{\partial t} \, . \label{el-eqs}
\end{equation}
It is easily checked that by the EL equations,
$(\dot{q}^i \partial \mathcal{L} / \partial \dot{q}^i -
\mathcal{L} )^{^.} = \partial \mathcal{L} / \partial t$.
Assuming that there is no explicit time dependence in the Lagrangian,
$(\dot{q}^i\partial{{\mathcal{L}}}/\partial\dot{q}^i-{\mathcal{L}})$ is a
conserved quantity, which is the energy, and is called the {\it Hamiltonian}.

So far only a system of particles has been considered. In earlier times a fluid was generally regarded as distinct from particles, though people like Democritus (c. 460-370 BC) argued that even water must finally be particulate. With Robert Boyle (1627-1691 AD) and John Dalton (1766-1844 AD) the idea was lifted out of the realms of Philosophy to a scientific base in Chemistry. Later, Daniel Bernoulli \cite{db} treated the flow of fluids mathematically as a nearly infinite system of particles. The term ``molecule" was coined much later by Amedeo Avogadro \cite{aa} and defines our modern view of material fluids. He (and his brother Johann, who claimed that he had written the work prior to Daniel) sent their work to Leonhard Euler.

As the third term comes from the explicit time-dependence of the Lagrangian, if it is positive it corresponds to dissipation of energy in Physics, such as the velocity dependent friction. For the Universe as a whole, i.e. in the cosmological context, it corresponds to the energy getting absorbed into the expanding spacetimes, like water into an expanding sponge. In Economics it would be the money ``evaporating" by inflation. To the contrary, if it is negative, it corresponds to absorption of energy from the environment in Physics and to energy getting ``squeezed out" of the spacetime, as from a sponge. In Economics it corresponds to deflation. If the analogy holds, we can expect the cosmological deflation to ``squeeze out" thermal radiation. Instead of merely being crushed to death, the Universe would be broiled and crushed to death.

\section{The Variational Principle for Fields}

\noindent Even if all material objects, including fluids like water and air, are made of particles, there are so many particles that one might as well take them to be infinitely many. Even that would not provide us with the full power of calculus, so it was worthwhile to use the continuum limit. Further, with Michael Faraday \cite{mf} and James Clerk Maxwell's \cite{jcm} theory of electricity and magnetism, they had to be treated as continuous fields. As such, it became necessary to extend the Lagrange formalism to a continuum in space and in time. Faraday and Maxwell's electric and magnetic fields, ${\bf E} ( t, {\bf x})$ and ${\bf H} ( t, {\bf x})$, pervade all of space at all times. Let us briefly review the developments.

Faraday's law relates changes in the magnetic field over time to changes in the electric field over space. Maxwell believed that the converse effect should also hold and modified Ampere's law relating the current density, ${\bf j}$ to a magnetic field varying over space, to include a change of the electric field over time. The Gauss equations use his divergence theorem to relate the divergence of the electric and magnetic fields to the strengths of their sources. Since there is no ``magnetic charge", the divergence of the magnetic field is zero. Since electric causes have magnetic effects and vice versa, Maxwell unified the theories of electricity and magnetism to electromagnetism. The
theory was formulated mathematically in a set of four PDEs, written in the cumbersome formalism of the time. In more modern notation they are:
\begin{eqnarray}
  \nabla \cdot {\bf E} & = & \frac{\rho}{\epsilon}   \qquad \qquad \qquad \qquad \qquad \qquad \, {\rm ( Gauss's \,\, law)}; \\ \nabla \cdot {\bf H} & = & 0  \qquad \qquad \qquad \qquad \qquad \qquad \,\, {\rm ( Gauss's \,\, law)};  \\ \nabla \times {\bf H} & = & \epsilon \frac{\partial {\bf E}}{\partial t} + {\bf j} \qquad \qquad {\rm ( Modified \,\, Ampere's \,\, law)}; \\ \nabla \times {\bf E} & = & - \mu \frac{\partial {\bf H}}{\partial t} + {\bf j} \qquad \qquad \qquad \quad {\rm ( Faraday's \,\, law)};
\end{eqnarray}
in electrostatic units, where $\rho$ is the charge density, $\epsilon$ is the dielectric constant of the material and $\mu$ its magnetic susceptibility (see, for example \cite{aq2}). The vacuum has a dielectric constant and magnetic susceptibility, denoted by a subscript zero. Thus the spacetime that carries the field takes the place of the aether - not a mechanical aether but an electromagnetic aether - which supports the field.

Notice the symmetry in these equations, in that two deal only with fields and two have material sources involved (the charge and current density). Even in the absence of these sources at some point in space at some time, the fields can still exist. There is also a symmetry between space and time in the equations. However, in Maxwell's theory there is a scalar potential, $\phi$ for the electric force and a vector potential, ${\bf A}$, for the magnetic field, so that ${\bf E} = - \nabla \phi$ and $\mu \, {\bf H} = \nabla \times {\bf A}$. While we can restore the symmetry of the Maxwell equations by considering the source-free case, how can we obtain a symmetry in the potentials? In his seminal paper on Special Relativity, Albert Einstein \cite{ae1} brought out the symmetry by unifying space and time into a single {\it spacetime}, so that a point in spacetime is given by a {\it four-vector}, $x^{\mu} = ( x^0, x^i) = ( ct,{\bf x})$. In the same way the electromagnetic potential is a four vector, $A^{\mu} = ( \phi / c, -{\bf A})$. Though the first field theory was for fluids, that was artificial, as it dealt with discrete particles {\it as if} they formed a continuum. In this case the fields {\it are} continua. In four-vectors we can write the Maxwell fields in terms of a tensor, $F_{\mu \nu} = A_{\nu,\mu} - A_{\mu,\nu}$ or,  in the language of forms ${\bf F} = {\bf d} \wedge {\bf A}$. The
corresponding Maxwell equations are
\begin{equation}
  F^{\mu \nu}_{\,\,\, ;\nu} = j^{\mu}\, , \qquad F_{\mu \nu , \rho} + F_{\rho \mu , \nu} + F_{\nu \rho, \mu} = 0 \, ; \label{maxwell-eqs}
\end{equation}
\begin{equation}
  {\bf d} \cdot {\bf F} = {\bf j} \, , \qquad {\bf d} \wedge {\bf F} = 0 \, .
\end{equation}
In fact, inserting the definition of the Maxwell field as the skew derivative (generalized curl) of the four-vector potential, the second equations are easily seen to be identities, ${\bf d} \wedge {\bf d} \wedge {\bf A} = 0$, as the exterior derivative, ${\bf d} \wedge$, is associative. In other words the definition of the four-vector potential makes the magnetic Gauss law and Faraday's law into identities, while the physical content now resides in the equations with sources: Gauss' electric law, and Ampere's modified law. The difference between ``$;$'' and ``$,$'' is explained in the next section.

It is easily shown (see for example \cite{aq1}) that in the absence of any source the electric and magnetic fields satisfy the wave equation and the speed of the electromagnetic wave, $1/\sqrt{\epsilon \mu}$, so that the speed of light in vacuum is $c = 1/\sqrt{\epsilon_0 \mu_0}$. Since the dielectric constant and magnetic susceptibility of the vacuum are less than for any material medium, the speed of the electromagnetic wave in vacuum is the maximum speed of these waves. Maxwell had already noted in his work \cite{jcm}, that this is the speed of light. Einstein had pointed out that it is the maximum attainable speed for {\it any} form of matter or energy \cite{ae1,ae2,ae3,ae4,ae5}.

We need to repeat the variational procedure for fields in $n$-dimensional spaces and four-dimensional spacetime, as slightly new features arise both times. Let us keep in mind three-dimensional spaces first, but be ready to extend the three to $n$. A scalar field, $\phi$, is an explicit function of
the independent variable, $t$ and of the three spatial variables $x^i(t)$. Though this is not necessary, we limit the Lagrangian to be a function of the field and its first total $t$-derivative,
${\mathcal{L}}[\phi(t,x^i(t)),\dot{\phi}]$, where $\dot{\phi}=\phi_t+\dot{x}^i\phi_i$. The problem now is what is meant by the derivative of the functional ${\mathcal{L}}$ with respect to a {\it function}
$\phi$, rather than a continuous variable like $t$? For the latter we just take the limit as $\delta t$ tends to $0$. A function can tend to zero in infinitely many ways. As such, we need a measure of the function, let us say the root mean square norm, and then let {\it that} tend to zero. To distinguish between the two concepts of derivative, we use $\delta$ in place of $\partial$. Now repeating the previous variational procedure, yields the EL equations for scalar fields
\begin{equation}
\frac{\delta{\mathcal{L}}}{\delta\phi}=\frac{\partial}{\partial t}
\left(\frac{\delta{\mathcal{L}}}{\delta\phi_t}\right)+
\left(\frac{\delta{\mathcal{L}}}{\delta\phi_i}\right)_{;i}~ \, ,
\end{equation}
where $\phi_t$ and $\phi_i$ stand for the partial derivative of $\phi$ relative to $t$ and $x^i$, respectively. For use in Mechanics we can take $i=1,2,3$, but in Economics we have to allow for all the commodities. It would be worth exploring the use of fields in Economics. If, instead of a scalar field, there is a vector field, $A_r(t, x^i(t))$, the EL equations become
\begin{equation}
\frac{\delta{\mathcal{L}}}{\delta A_r}=\frac{\partial}{\partial t}
\left(\frac{\delta{\mathcal{L}}}{\delta A_{r,t}}\right)+
\left(\frac{\delta{\mathcal{L}}}{\delta A_{r,i}}\right)_{;i}~ \, ,
\end{equation}
You might have expected that the examples of the electric and magnetic fields would be given here, but the fact is that a field theory of one without the other would be incomplete. We need to use the relativistic electromagnetic field, $A_{\mu}$. However, there is a complication. In relativity, time is like any other coordinate, so we do not have a separate independent variable, but four independent variables,  and the field is $A_{\mu}(x^{\nu})$. Now the variation is with respect to each function of each variable. The EL equations for a relativistic vector field become,
\begin{equation}
\frac{\delta{\mathcal{L}}}{\delta A_{\mu}}=\left(\frac{\delta{\mathcal{L}}}
{\delta A_{\mu,\nu}}\right)_{;\nu}~. \label{eq20}
\end{equation}
Using the electromagnetic Lagrangian,
$${\mathcal{L}}=A_{\mu}j^{\mu}+\frac{1}{4}F_{\mu\nu}F^{\mu\nu},$$
in the above EL equations gives the first of the Maxwell equations \eqref{maxwell-eqs}. The other is, of course, an identity.

\bigskip
\section{A Geometrical Application of the Lagrangian}
\label{lagr-geodesics}

\noindent  The geometrical Lagrangian is obviously for a continuum, but not
so obviously for a field theory. If we limit our discussion to
three-dimensional space, or surfaces, or even to Minkowski space, it is {\it
not} for a field. As we saw, electromagnetism, which is the epitome of a
field theory, has a vector field, $A_{\mu}$. Geometrically, the arc length
square is given by $$ds^2=g_{ij}(x^k) dx^i dx^j,$$ where $g_{ij}$ is the
matrix representation of the metric tensor, ${\bf g}$, in some
$n$-dimensional coordinate system in index notation and $dx^i$ are
infinitesimal changes of the position vector in those coordinates \cite{aq2}.
Thus geometry needs a (second rank) tensor field, $g_{ij}$. Note that not
only the metric coefficients, but the metric tensor itself, varies from point
to point.

The shortest path between two points, P and Q, called a {\it geodesic}, is
obtained by minimizing the integral of the arc length, $ds$, along the path
from one to the other,
\begin{equation}
s_{PQ}= \int_P^Q ds = \int_P^Q \ell [x^i(s),\dot{x}^j(s)] ds =
\int_P^Q g_{ij}(x^k)\dot{x}^i\dot{x}^j ds~ \, ,
\end{equation}
where $\ell [x^i(s),\dot{x}^j(s)]$ is the Lagrangian, which has a constant
value as a{\it function} of $s$, but as a {\it functional} it depends on the
position and velocity vectors. Using the EL equations, \eqref{el-eqs},
without explicit dependence on the parameter $s$, we obtain
\begin{equation}
\frac{d}{ds}(g_{ij}\delta^i_k\dot{x}^j+g_{ij}\delta^j_k\dot{x}^i)=
g_{ij,k}\dot{x}^i\dot{x}^j~.
\end{equation}
The total derivative of $\dot{x}^i$ is $\ddot{x}^i$ and, since $g_{ij}$ is an
explicit function of $x^i$ but not of $s$,
$\frac{d}{ds}g_{ij}=g_{ij,l}\dot{x}^l$. Hence
\begin{equation}
g_{kj} \ddot{x}^j + g_{ik}\ddot{x}^i + g_{kj,l}\dot{x}^j\dot{x}^l +
g_{ik,l} \dot{x}^i \dot{x}^l = g_{ij,k} \dot{x}^i \dot{x}^j~. \label{eq23}
\end{equation}
As the metric tensor defines the length of a vector, so it and its inverse
must exist at every point. In index notation we write the inverse as $g^{ik}$
such that $g_{ij}g^{ik}=\delta^j_k$, the Kronecker delta, which is the
identity matrix in index notation. Multiplying \eqref{eq23} through by half
the inverse metric, relabeling dummy indices and transposing the term on the
right side, we obtain the {\it geodesic equation}
\begin{equation}
\ddot{x}^i + \Gamma^i_{jk} \dot{x}^j \dot{x}^k = 0~ \, , \label{geo-eqn1}
\end{equation}
where $\Gamma^i_{jk}$ is the {\it Christoffel symbol}, defined by
\begin{equation}
\Gamma^i_{jk} = \frac{1}{2} g^{il} \left( g_{jl,k} + g_{kl,j} -
g_{jk,l} \right)~ \, , \label{chris}
\end{equation}
which gives the difference between the covariant derivative denoted by
``$;k$'' and the partial derivative denoted by ``$,k$'', namely for any
$A^i$,
\begin{equation}
A^i_{;k} = A^i_{,k} + \Gamma^i_{jk}A^j~.
\end{equation}

One of Euclid's theorems for a plane says that, the shortest path between two
points is the straight line joining them. Of course, there is no straight
line in a curved space (like the surface of the Earth, where straight lines
in three dimensions are excluded). The {\it straightest} available path in an
$n$-dimensional space, is the curve whose unit tangent vector, ${\bf t}$,
does not change direction along it, i.e. $\frac{d}{ds}{\bf t}=0$. Now, we
could write the tangent vector in the local coordinates discussed, with the
basis vectors, ${\bf e}_i$, so that ${\bf t}=\dot{x}^i{\bf e}_i$. Since every
vector can be written as a linear combination of basis vectors, so the
partial derivative along along any direction can be so written. Hence, ${\bf
e}_{i;j} = \Gamma^k_{ij}{\bf e}_k~$, As such the Christoffel symbol arises
from the differentiation of the basis vectors. It is shown in Differential
Geometry (see for example \cite{aq2}) that this set of coefficients is given
by \eqref{chris}. Thus the generalization of Euclid's theorem to curved
spaces is: ``the shortest {\it available} path between two points, is the
straightest {\it available} curve.'' In a flat space in Cartesian coordinates
the basis vectors are constant, but in general, like in Gauss' theory of
surfaces, they vary. The simplest examples are of the polar basis vector in a
plane in polar coordinates, and all basis vectors on a sphere.

\section{Symmetries of Fields and Lagrangians}

\noindent What would be meant by the symmetry of a field? For one thing, the
field may not depend on some independent variable(s), such as time or
position. Writing the field, which may be a scalar, vector or tensor, as
${\bf A}$, then $\partial{\bf A}/\partial t = 0$, or the equivalent for one
or more position variable. In that case we say that ${\bf
X}=\partial/\partial t$, or the equivalent for some position variable is a
{\it translation symmetry} of the field. On the other hand, it could be that
the field itself ``inflates'' (or ``deflates'') with time, or along some
spatial direction. In that case $t\partial/\partial t$ or $x\partial/\partial
x$ say, is called a {\it scaling symmetry} of the field. Even if the field
depends on the independent variables, it could be that some physically (or
economically) relevant quantities do not. For example it may be that the
energy and momentum re-scale in such a way that the difference between the
square of the energy and a constant times the square of the other is
constant, as in Special Relativity, where $E^2-p^2c^2=m^2c^4$. In this case
we say that the mass, $m$, is an {\it invariant} and the corresponding
infinitesimal symmetry generator is $p^{\mu}=\partial/\partial x^{\mu}$,
which gives the energy momentum four-vector. Notice that this says that
energy-momentum is collectively but not separately, conserved. The conserved
quantity is the Hamiltonian, $m c^2$. In Physics one would call the former
conservation a ``conservation law'', but people in Symmetry Analysis call the
latter by that name. There can be other symmetry generators, like a combined
scaling symmetry such as $x\partial/\partial x+y\partial/\partial y$, along
the line $y = x$, or rotations given for example by $y\partial/\partial
x-x\partial/\partial y$ for rotation about the $z-$axis. In Relativity, the
Lorentz transformation for uniform linear motion in the $x$-direction is
$ct\partial/\partial x+x\partial/\partial(ct)$. Thus in Special Relativity
this would be a conservation law. There can also be symmetries of
combinations of the field, or of it and the first derivative of the field,
but not of the field itself.

Of special relevance are symmetries of the Lagrangian, because it gives the
dynamics arising from the variational principle for the Lagrangian, while
also allowing a reduction of the number of variables that are involved in the
DE, or reducing its order. This double reduction makes these symmetries
especially useful. They are called {\it Noether symmetries}. Noether's
theorem says that to each symmetry of the Lagrangian, there corresponds a
conserved scalar quantity (called a {\it Noether charge}). Each conserved
quantity is a first integral of some part of the equations of motion. At the
same time, when we use that quantity as a new ``variable'', it reduces the
equation by trivializing that part of it. The most obvious invariant is the
Lagrangian itself by definition. This is not useful for any reduction, as it
is tautologically true. In fact, scaling it by a constant cannot change the
equations of motion, since they are linear and homogeneous in the Lagrangian.
The great thing about using Geometry for kinematics or dynamics is that in
that case every geometrical symmetry will be a non-trivial Noether symmetry
and provide a double reduction for the equations of the theory. The charges
give us physical information and help to reduce the order of the equations.
Before continuing with the applications, we need to present the definition of
Noether symmetries and explain how they are determined.


\subsection{ Noether Symmetries }
\label{noether-sym}

\noindent Noether's theorem \cite{noether} is applicable to
a dynamical system of ordinary or partial differential equations obtained
from a variational principle \cite{olver,bluman-anco,ibragimov,bluman2010}.
Let $x^a$ be $\ell$ independent variables and $q^i$ be $m$ dependent
variables which are arbitrary (sufficiently smooth) functions of independent
variables. The total derivative operator $D_x$ given in \eqref{eq5} can be
recast into the form
\begin{equation}
D_a = \frac{\partial}{\partial x^a} + q^i_a \frac{\partial}{\partial \,
q^i } + q^i_{a b} \frac{\partial}{\partial \, q^i_b } + ... \, , \label{eq5-2}
\end{equation}
where the derivatives of $q^i$ with respect to $x^a$ are represented by $
q^i_a = D_a q^i\, , q^i_{a b} = D_b D_a q^i$, and so on. Then, the
Euler-Lagrange operator, for each $i$, is defined by
\begin{equation}
\frac{\delta}{\delta q^i } = \frac{\partial}{\partial q^i } + \sum_{N \geq 1}
\left( -1 \right)^N D_{a_1} ... D_{a_N} \frac{\partial}{\partial \,
q^i_{a_1 ..a_N } }\, . \label{euler-lagr-opr}
\end{equation}
We now consider the EL equations of motion
\begin{equation}
  U^k \left( x^a, q^i, q^i_{(1)}, ..., q^i_{(N)} \right)  = 0 \, ,
  \label{euler-lagr}
\end{equation}
which is an ${\it N}^{th}$-order system of $m (\geq 1 )$ PDEs or, the ODEs if
$\ell=1$. Eq. \eqref{euler-lagr} is assumed to be of maximal rank and locally
solvable, where the collection of {\it N}th-order derivatives is denoted by
$q^{i}_{(N)}$. If there exists a function $\mathcal{L} ( x^a, q^i, ...,
q^{i}_{(M)})$, \, $M < N$, such that \eqref{euler-lagr} is equivalent to
\begin{equation}
  \frac{\delta  \mathcal{L} }{\delta q^i } = 0 \, , \label{euler-lagr-2}
\end{equation}
then $\mathcal{L}$ is called a Lagrangian of \eqref{euler-lagr}. Here the
${\it M}^{th}$ prolonged operator has the form \cite{ibr1998}
\begin{eqnarray}
{\bf X}^{[M]} = {\bf X} +  \eta'^{i}_{\, a}  \frac{\partial}{\partial \,
q^{i}_a } + \eta'^{i}_{\, a b}  \frac{\partial}{\partial \, q^{i}_{a b} }+
...  + \eta'^{i}_{\, a_1 ... a_M}  \frac{\partial}{\partial \,
q^{i}_{a_1 ... a_{M}} } \, ,
\end{eqnarray}
where the operator ${\bf X}$ is of the form
\begin{equation}
  {\bf X} = \xi^a (x^b, q^{j}) \frac{\partial}{\partial x^a} +
  \eta^{i} (x^b, q^{j}) \frac{\partial}{\partial \, q^{i} } \, ,
  \label{noether-opr}
\end{equation}
and $\eta'^{i}_{\, a}$ and $\eta'^{i}_{\, a b}$ are defined by
\begin{equation}
  \eta'^{i}_{\, a} = D_ a \eta^{i} - q^{i}_b D_a \xi^b \, ,
  \quad \eta'^{i}_{\, a b} = D_ a D_b \eta^{i} - q^{i}_{b c} D_a \xi^c -
  q^{i}_{a c} D_b \xi^c - q^{i}_c D_a D_b \xi^c \,  .
\end{equation}

A Noether symmetry generator corresponding to a Lagrangian
$\mathcal{L} ( x^a, q^{i}, ..., q^{i}_{(M)})$ is the operator ${\bf X}$ in
\eqref{noether-opr} if there exists a vector ${\bf K} = ( K^1, ...,
K^{\ell})$, or a function $K (x^a, q^{i} )$ if there is only one independent
variable (i.e., $\ell=1$), such that
\begin{equation}
{\bf X}^{[M]} (\mathcal{L}) + \mathcal{L} \, D_a ( \xi^a ) = D_a ( K^a )\, ,
\end{equation}
where $M < N$.
Further, the operator ${\bf X}^{[M]}$, which is also called the
Lie-B\"{a}cklund symmetry, is a Noether symmetry of $\mathcal{L}$
corresponding to an Euler-Lagrange equations \eqref{euler-lagr-2} if and only
if the Lie characteristic function $W^i = \eta^{i} - \xi^b \, q^{i}_b $ of
${\bf X}^{[M]}$ is also the characteristic of the conservation law
\begin{equation}
  D_a ( I^a ) = 0 \, , \label{consv-law}
\end{equation}
where $I^a$ has the form
\begin{equation}
  I^a = N^a ( \mathcal{L} ) - K^a \, , \label{consv-formula}
\end{equation}
and the Noether operator associated with the operator ${\bf X}^{[M]}$ is
defined by Ibragimov \cite{ibr1979} as follows:
\begin{eqnarray}
N^a = \xi^a + W^{i} \frac{\delta}{\delta \, q^{i}_a } + \sum_{M \geq 1}
D_{a_1} D_{a_2} ... D_{a_{M}} \left( W^{i} \right) \frac{\partial}{\partial
\, q^{i}_{a \, a_1 ... a_{M}} }  \, .
\end{eqnarray}
Here ${\bf I} = (I^1,...,I^{\ell})$ will be called a {\it conserved vector}
of the EL equations \eqref{euler-lagr}, or a {\it conserved quantity} (or
{\it first integral}) if $\ell=1$. Notice that for variational problems with
Lagrangian functions depending on higher-order derivatives, the main
conservation theorems are valid, but the conserved vector
\eqref{consv-formula} for a Lagrangian function depending on any order
derivatives has a different form.

\subsection[Symmetries of Fields and Lagrangians]{Classical Mechanics}

\noindent Time-translational invariance of the Lagrangian in Classical
Mechanics reduces the number of variables in the equations of motion from
four to three and it implies energy conservation. The former property makes
it easier to solve the ODEs involved, but the latter allows one to {\it know
the answer without solving the equations}. For a simple harmonic oscillator,
one can use the further spatial symmetry to reduce them to a single
second-order ODE with constant coefficients, but the momentum conservation
gives planar motion and the law of energy conservation says that a
frictionless oscillator will continue its motion forever, and that the
greater the friction the faster the motion will die out. The formal aspects
of the calculations of Noether symmetries in this case are explained below.

Consider a first order Lagrangian for only one
independent variable, so that the EL equations of motion are given by
\eqref{el-eqs}. Then the energy functional associated with $\mathcal{L}$ is
defined by
\begin{equation}
E_{\mathcal{L}} = { \dot{q} }^{i} \frac{\partial \mathcal{L}}{\partial
{ \dot{q} }^{i} } - \mathcal{L}, \label{energy}
\end{equation}
which is also the Hamiltonian of the system. This yields the first integral
as a system of ODEs of the form
\begin{equation}
{ \ddot{q} }^{i} = w^i (\tau, q^{k}, {\dot{q}}^{k} ). \label{ode}
\end{equation}
The Noether symmetry generator for this Lagrangian is given by
\begin{equation}
  {\bf X} = \xi (t, q^{i}) \frac{\partial}{\partial t} + \eta^{i} (t, q^{i})
  \frac{\partial}{\partial \, q^{i} } \, , \label{noether-first-opr}
\end{equation}
if there exists a function $K (t,q^{k} )$ and the Noether symmetry condition
\begin{equation}
{\bf X}^{[1]} ( \mathcal{L} ) + \mathcal{L} \,  D_{t} ( \xi )=
D_{t}  K  \, , \label{ngs-cond-first}
\end{equation}
is satisfied. Here ${\bf X}^{[1]}$ is the first prolongation of Noether
symmetry generator ${\bf X}$, i.e.,
\begin{eqnarray}
& & {\bf X}^{[1]} = {\bf X} + {\eta'}^{i} (t, q^{k}, { \dot{q} }^{k} )
\frac{\partial}{ \partial  { \dot{q} }^{i} } \, , \label{first-pro}
\end{eqnarray}
where ${\eta'}^{i} (t, q^{k}, { \dot{q} }^{k} ) = D_{t} \eta^{i} - { \dot{q}
}^{i} D_{t} \xi$. For every Noether symmetry generator ${\bf X}$, the
conservation law \eqref{consv-formula} becomes $D_{\tau} ( I ) = 0$, where
$I$ is the corresponding Noether flow, and it has the expression
\begin{equation}
I = - \xi E_{\mathcal{L}} + \eta^{i} \frac{\partial \mathcal{L}}{\partial
{ \dot{q} }^{i} } - K \, , \label{first-int}
\end{equation}
which is a {\it conserved quantity} of the system of equations \eqref{ode}.

There is a second way of finding symmetries of a Lagrangian
$\mathcal{L}$ for a given dynamical system, called the {\it strict Noether
symmetry approach} that yields $\pounds_{\bf X} \mathcal{L} = 0$, where
$\pounds_{\bf X}$ is the Lie derivative operator along ${\bf X}$
\cite{capo1996}. The only difference is that here $K$ vanishes. Both
approaches are useful in a variety of problems arising from physics and
applied mathematics, and lead to first integrals. The important thing is that
which approach yields a conserved quantity. The classical Noether symmetries
have the advantage of yielding conserved quantities or conservation laws,
directly \cite{ibragimov}. Cyclic variables are also used and are related to
Noether symmetries, but there is ambiguity in their choice. For more details
see \cite{capo2007}.

Another example is of a particle that would normally undergo
geodesic motion, but is forced off it, like a charged particle in an EM
field. For the present purpose take a scalar potential  $V(s,x^{k})$, then the
geodesic Lagrangian describing the motion of the massive or massless (i.e.,
lightlike)  particles  can be written as
\begin{equation}
\mathcal{L} (s, x^k, \dot{x}^k) = \frac{1}{2} g_{i j} \dot{x}^i \dot{x}^j -
V(s,x^k) \, ,  \label{geodesic-lagr}
\end{equation}
which gives rise to the forced geodesic equations of motion
\begin{equation}
\ddot{x}^i  + \Gamma^i_{j k} \dot{x}^j \dot{x}^k =  F^i ~ \, , \label{geo-eqn2}
\end{equation}
where $F^i = g^{i j} V_{,j}$ is the conservative force field. For every
Noether symmetry, there is a {\it a first integral} for the system of
equations \eqref{geo-eqn2} of the form
\begin{equation}
I=- \xi E_{\mathcal{L}}+ g_{i j} \eta^{i} \dot{x}^{j} - K \, , \label{first-int-geo}
\end{equation}
where the {\it energy functional} \eqref{energy} for the geodesic Lagrangian
is given by
\begin{equation}
E_{\mathcal{L}} = \frac{1}{2} g_{i j} \dot{x}^{i} \dot{x}^{j} + V(s, x^{k}) \, ,
\label{energy-geo}
\end{equation}
which is the Hamiltonian of the system.

\subsection[Symmetries of Fields and Lagrangians]{Economics}

\noindent Shifting from Mechanics to Economics, time-translational invariance of the Lagrangian implies that prices will stay constant over time and the total amount in the Economy stays constant. ``Amount of what?'', you ask. This hides an aspect of Economics with no Classical Mechanics analogue. Two
distinct quantities may be taken: wealth; or money. Wealth refers to the {\it actual} goods and services in the Economy, while money is an {\it arbitrary} measure for that wealth, called a ``numeraire''. Being arbitrary, we can change the amount of wealth to a unit of money. The Government may do this to count in wealth that would be there in the Economy, but not as yet registered in the accounting; or to appear to be richer. In the former case, there is a growing economy, and the ``value of money'' remains constant. In the latter case the value of money will decline while the Economy remains stagnant. The Government could even anticipate the wealth yet to be generated, before it has been, so as to accelerate economic growth. As such, it would ``borrow from the future''. This is called ``credit creation''. It can lead to runaway inflation leading to a credit crunch, as has been seen.

Already we have seen economic insights provided by the invariant beyond the benefit of reducing the number of variables in the EL-equations. More follow. In 1945 John von Neumann demonstrated \cite{jvn} that the rate of interest equals the rate of growth for an optimally growing Economy. In other words,
for optimal growth, the amount of money in the Economy must grow at the same rate as the amount of wealth. It was then shown that \cite{aq3,aq32} that there must be inflation in a growing economy. The excess money behaves in much the same way as the entropy in Physics corresponding to the shortfall of efficiency of a heat engine. It should be noted that the Noether invariant in both applications (Physics and Economics) has significance {\it far beyond} its use for reducing the number of variables and order of the governing ODEs.

Though there is no analogue of the two ways of considering the ``amount'' of something in {\it Classical} Mechanics, there is in {\it Relativistic} Mechanics. The measures of length and time vary from observer to observer, since they are arbitrary measures of the invariant length and duration. Consequently, the invariants play a far more significant role in Relativity than they do in Classical Mechanics. To explain this will need to provide a quick review of geometrical symmetries.

\subsection[Symmetries of Fields and Lagrangians]{Geometrical Symmetries}

\noindent There are two generalizations of the derivative for a manifold. One is to first map an open set on the manifold to a coordinate frame, $\mathbb{R}^n$, take the derivative along a vector in the usual way and then map the derivative back to the manifold. The second is to directly take the rate of change by
infinitesimal movement along a curve on the manifold. The former is called the {\it intrinsic derivative} and the latter the {\it Lie derivative}. The former procedure obviously incorporates the derivatives of the basis vectors, while the latter does not. The intrinsic derivative, left in the coordinate system, is called the covariant derivative and denoted by ``$;$'', as we saw.
The Lie derivative along a vector field, ${\bf t}$, will be denoted by $\pounds_{\bf t}$. If the intrinsic derivative is used in a Taylor series to move from one point on the manifold to another, it is called {\it parallel transport}, as a curve traced out by transporting a vector field, ${\bf p}$, along a curve with tangent vector field ${\bf t}$, will appear parallel as seen in the coordinate system. However, it will not be parallel on the manifold. If the Lie derivative is used for moving on the manifold it is called {\it Lie transport}.

This is most easily seen by taking a sphere as the manifold and the base curve to be a line of latitude. Transporting a unit North pointing vector traces out the next line of latitude. At the equator it is not so obvious, but near the North pole, as we know from our school Geography, on the map of the Earth it does not look parallel, and the square on the map does not look like a square on the globe. The reason is that the {\it lengths} of the upper and lower sides of the ``square'' on the globe are unequal, but the {\it angles} subtended by them are the same. Thus if ${\bf p}$ takes point $P$ to point $Q$ on one line of latitude, and ${\bf t}$ takes $P$ to $R$ on the same line of longitude and $Q$ to $S$ on the next one, by definition of subtending the same angle, ${\bf p}$ must take $R$ to $S$, so ${\bf p}$ is Lie transported along ${\bf t}$ and the square on the globe closes. However, since the lengths of the upper and lower ${\bf p}$ are unequal, it cannot close on the map.

Let us get a bit more technical. If ${\bf p}$ is to be Lie transported from $P$ to $Q$ by ${\bf t}$, then ${\bf p}|_Q=exp(\pounds_{\bf t}){\bf p}|_P$. Now, if ${\bf p}$ is {\it invariant} under this Lie transport, then clearly $\pounds_{\bf t}{\bf p} = 0$. In this case the ``square'' mentioned above, or more generally the ``rectangle'', closes and so ${\bf t}{\bf p}={\bf p}{\bf t}$. Taken into the coordinate system, $t^a p^b_{;a} = p^a t^b_{;a}$. Now, since the Christoffel symbol is symmetric the terms involving it on both sides, $p^at^c\Gamma^b_{ac}$ and $t^ap^c\Gamma^b_{ac}$, cancel and so we have $t^ap^b_{,a} = p^a t^b_{,a}$. {\it This} is the way that the derivative of the basis vector is removed from the Lie derivative.

\subsection[Symmetries of Fields and Lagrangians]{General Relativity}

\noindent First let us introduce General Relativity (GR). It requires that
all observers be equally good. ``Observers'' are conceived as disembodied
persons --- massless points endowed with a clock attached to a spatial frame
of reference --- who move on geodesics. ``Equally good'' means that physical
laws are no simpler in one frame than another. This does not necessarily mean
that one cannot distinguish between different frames. For example, one could
feel acceleration, so a frame of zero acceleration could be determined, but
the presence or absence of acceleration would not be relevant for the
statement of a physical law. As such, those laws must be defined on
manifolds. Special Relativity (SR) {\it does} assume that there is no way to
distinguish between two un-accelerated frames. However, this must again be
seen in a special context of two observers communicating with each other,
without reference to a third \cite{aq2}. By choosing the frame of reference
in which the cosmic microwave radiation is isotropic (up to statistical
fluctuations) we {\it can} determine the rest-frame of the Universe. As such,
GR entails using Lie derivatives and Lie transport.

As in the earlier example, let us take ${\bf t}$ to be the unit tangent
vector to the geodesic of an observer, $O$ and ${\bf p}$ be the position
vector of another observer $O^{\prime}$ relative to $O$. Then ${\bf t}({\bf
p})=\dot{{\bf p}}$ and ${\bf t}[{\bf t}({\bf p})]=\ddot{{\bf p}}$, i.e. the
velocity and the acceleration of $O^{\prime}$ as seen by $O$. In Geometry,
the latter is called {\it geodesic deviation} and will be denoted by ${\bf
\mathcal{A}}$. In index notation
\begin{equation}
{\mathcal{A}}^{\mu}=t^{\alpha}(t^{\nu}p^{\mu}_{;\nu})_{;\alpha}~.
\end{equation}
Using the Lie transport requirement we can interchange the $t$ and $p$ inside
the bracket and, on expanding the bracket use the geodesic condition on one
term. What remains is
\begin{equation}
{\mathcal{A}}^{\mu}=t^{\alpha}t^{\nu}[p^{\mu}_{;\nu;\alpha}-p^{\mu}_{;\alpha ;\nu}]~.
\end{equation}
Now, by definition, the Riemann curvature tensor is defined \cite{mtw} by the
skewed second derivative of a vector, so that we get
\begin{equation}
{\mathcal{A}}^{\mu}=R^{\mu}_{~\alpha \nu\beta}t^{\nu}p^{\alpha}t^{\beta}~.
\end{equation}
This means that acceleration is related to the curvature of the spacetime
manifold. Physically, acceleration is caused by matter and energy. We saw how
energy conservation arose for a system of particles in Classical Mechanics.
It would be useful to generalize that concept to other physical fields in a
four-dimensional spacetime.

On account of mass-energy equivalence we no longer have conservation of mass
and energy separately, but only of the matter-energy tensor, also called the
stress-energy tensor, (see \cite{mtw}, Ch.4). It includes the momentum
4-vector and spatial fluid stress tensor $\sigma^{ij}=dF^i/dS_j$, where
$dF^i$ is the force acting on an area element of the fluid, $dS_j$. If it is
an irrotational perfect fluid, the stress energy tensor density is
\begin{equation}
T^{\mu\nu}=\rho c^2\delta^{\mu}_0\delta^{\nu}_0 + \sigma^{ij}\delta^{\mu}_i\delta^{\nu}_j~,
\end{equation}
and the usual conservation laws for relativistic fluids, derived by the
requirement that the flux of fluid and energy crossing a closed surface is
conserved. Using Gauss' divergence theorem, it gives
\begin{equation}
T^{\mu\nu}_{\quad ;\nu}=0~.
\end{equation}
For a field with a vector-valued potential $A_r$, for an irrotational perfect
fluid, the stress energy tensor density is
\begin{equation}
T^{\mu\nu} = g^{\mu\alpha} A_{r,\alpha} \frac{\delta{\mathcal{L}}}{\delta A_{r,\nu}}
- g^{\mu\nu}{\mathcal{L}}~. \label{eq32}
\end{equation}
For $r=1$ we get a scalar field, for $r=\mu$ we get a scalar field, and for
$r=\mu\nu$ we get a tensor field. For any number of fields, we simply have to
add the stress energy tensor for each one to get the total stress-energy
tensor.

Returning to GR, we need that ${\bf T}$ be related to the Riemann tensor,
${\bf R}$, by a second rank tensor function, which is divergence-free. Since
the function is to be second rank and the Riemann tensor is fourth rank, we
need to take the trace of the Riemann tensor, namely the Ricci tensor. Using
the Bianchi identities that are satisfied by the Riemann tensor, we obtain
the linear combination of the Ricci tensor and scalar that is
divergence-free, ${\mathcal{E}}_{\mu\nu}=R_{\mu\nu}-\frac{1}{2}Rg_{\mu\nu}$,
called the {\it Einstein tensor}, yielding the simplest non-trivial relation,
called the {\it Einstein Field Equations}, (EFEs)
\begin{equation}
\kappa T_{\mu\nu} = R_{\mu\nu}-\frac{1}{2}Rg_{\mu\nu}+\Lambda g_{\mu\nu}~, \label{efe}
\end{equation}
where $\kappa=8\pi G/c^4$, gives the coupling of gravity with matter, $G$
being Newton's constant, $\Lambda$ is a constant of integration, called the
``cosmological constant''. In their full generality they are a system of ten
non-homogeneous, second order nonlinear PDEs for ten functions (metric
coefficients) of four variables. Even given the source, it would be
impossible to solve them generally.

Let us now take a quick look at Noether symmetries for the geodesic Lagrangian \eqref{geodesic-lagr} in GR for some well-known spacetimes which have been used ofr classifiation according to their symmetry
generators. This has been done for static plane, static spherical, and static
cylindrically symmetric spacetimes by Feroze and her collaborators
\cite{feroze1,feroze2,feroze3,feroze4,feroze5,feroze6}. The complete
classification of non-static plane and non-static spherically symmetric
spacetimes via Noether symmetry worked by Jamil et al. \cite{jamil1,jamil2}.
The Lie and Noether symmetries of geodesic equations have been studied for the Friedmann metrics by Tsamparlis and Paliathanasis \cite{tsamparlis1}. These symmetries have also been obtained for some of the Bianchi-type spacetimes in \cite{tsamparlis2,ptm2015,ah2016,hy2017}.  The complete analysis of Noether symmetries for G\"{o}del-type and pp-wave spacetimes studied by Camci et al.
\cite{ug2014,ug2015,uy2015}.

\subsubsection{Spacetime Symmetries}

\noindent This is where the need to use symmetries comes in. Given enough symmetries we can reduce the number of independent variables. Thus, if we have the maximum possible symmetry of a flat space, i.e. Minkowski space, the functions are fully given directly, being constant in Cartesian coordinates and only trivially dependent on the independent variables in other coordinates. This brings out a problem of determining whether the dependence is trivial or not. Can a change of the independent variables remove the apparent dependence on them? For this purpose we need an invariant characterization of the symmetry involved. Since the metric tensor is the (tensor) potential, we need that the symmetry direction be one along which ${\bf g}$ is Lie transported. If the symmetry vector is denoted by ${\bf k}$, we need that $\pounds_{\bf k}{\bf g} = 0$. In index notation this reduces to the {\it Killing equation},
\begin{equation}
k_{(\mu;\nu)} = 0 = k^{\alpha}_{,\mu}g_{\alpha\nu} + k^{\alpha}_{,\nu}g_{\alpha\mu} +
k^{\alpha} g_{\mu\nu,\alpha}~.
\end{equation}
A vector, ${\bf k}$, satisfying this equation is called a {\it Killing vector} (KV), or an {\it isometry}. If ${\bf k}$ is a unit timelike isometry then we can choose coordinates such that $k^{\mu}=\delta^{\mu}_0$ and use it to define the time coordinate and the metric coefficients will be time-independent in these coordinates. Similarly, if there are three unit spacelike isometries, ${\bf k}_i$, we can use them to define three spatial coordinates by $k^{\mu} = \delta^{\mu}_i$, for
$(i=1,2,3)$. In the former case the metric coefficients will be independent of time, and in the latter of space. As such, we will have time translation invariance in the former case and space translation invariance in the latter case. The Noether invariants corresponding to them will be the {\it energy}
and {\it momentum}. If the spacelike vectors generate the Lie algebra $SO(3)$, there can only be trivial dependence of the metric coefficients on the three coordinates and there will be rotational invariance. The corresponding conserved quantity will be the {\it angular momentum}. If the
timelike and spacelike vectors in pairs, generate an $SO(1,1)$ there will also be invariance under Lorentz transformations and the corresponding conserved quantity will be the {\it spin angular momentum}. If all (six) of these ``rotational'' symmetries exist, the Lie algebra will be $SO(1,3)$. If the four translations also exist the spacetime must be flat and one can choose Cartesian coordinates, in which the metric tensor is diagonal, being $1$ in the time component and $-1$ in each spatial component.

In a curved spacetime the symmetry cannot be higher dimensional, but can have the same number of dimensions if the curvature is a non-zero constant. Thus the number of symmetry generators will be ten and the associated group will be the {\it de Sitter} (dS) group, $SO(1,4)$ for positive curvature and the {\it anti-de Sitter}, (AdS) group, $SO(2,3)$, both of which contain $SO(1,3)$ as a subgroup. Since the metric has a timelike KV, there is still energy conservation. However, the distinction between linear and angular momentum disappears. The point is that the geodesic on which the linear momentum was being conserved, now bends around and closes on itself for dS and bends hyperbolically away for AdS, so that it becomes rotational motion as well. In all other cases some conservation laws will be lost.

To be concrete, the dS metric is:
\begin{equation}
ds^2= e^{\nu(r)} dt^2 - e^{-\nu(r)}dr^2 - r^2 \left( d\theta^2 + \sin^2\theta d\phi^2 \right)~ \, , \label{ds-metric}
\end{equation}
where distances are measured in light seconds and
\begin{equation}
e^{\nu(r)}= 1 - \frac{r^2}{R^2} \, , \label{ds-metric-nu1}
\end{equation}
$R$ being a constant. If, instead,
\begin{equation}
e^{\nu(r)} =  1 + \frac{r^2}{R^2}  ~ \, , \label{ds-metric-nu2}
\end{equation}
we have the AdS metric. In these cases the stress-energy tensor is proportional to the metric tensor and so, by the EFEs, the Ricci tensor must also be proportional to the metric tensor. Such spaces are called {\it Einstein spaces}. In the former case the stress-energy tensor has a positive trace, and in the latter a negative trace. Were we to interpret that trace as the energy, which would be done for a fluid, the former would have a positive energy and the latter a negative energy. The difference this makes may be best visualized by conceiving of a very large space city, in which some live on the outside and others on the inside. Those on the outside would see a horizon below which there is nothing, albeit smaller than they would see on Earth. The latter would see everything collecting up at the horizon and above that a sky that would look just like the city around them. The former is what
one gets with positive energy and the latter what arises with negative energy.

The gravitational field for a point particle of mass $m$ and charge $Q$ is given by
\begin{equation}
e^{\nu(r)}= 1 - \frac{2m}{c^2 r} + \frac{Q^2}{r^2} ~ \, ,
\end{equation}
in units with Newton's gravitational constant chosen to be unity, and yields a traceless Ricci tensor. If there is no charge, $Q=0$, the spacetime is Ricci-flat. The former is called the Reissner-Nordstr\"om metric, and the latter the Schwarzschild metric. In both cases we are only left with time translational and rotational invariance, or energy and angular momentum conservation. The fact that momentum conservation is lost is easily seen by considering a test particle left near a gravitational source, in which case it will fall towards the source. That spin angular momentum conservation is lost has the consequence that precession can be generated or lost in a
gravitational field, and can be tested. For gravitational waves time translation is also lost and so energy is not conserved.

Some more conservation laws are found in an {\it Einstein Universe}, which has the symmetry group, $SO(4) \bigotimes \mathbb{R}$, for which the coefficient of the time metric coefficient in Eq. \eqref{ds-metric}, is unity and the coefficient of the radial coefficient is given by Eq. \eqref{ds-metric-nu1}. This is an Einstein space and we recover the usual translation invariance as well, leading to linear momentum conservation. If Eq. \eqref{ds-metric-nu2} is satisfied instead of Eq. \eqref{ds-metric-nu1}, it is the {\it anti-Einstein Universe} and the usual angular momentum conservation is replaced by a spin angular momentum conservation, as we get an $SO(1,2)$ instead of an $SO(3)$, symmetry group. The total number of conserved quantities is seven.

There are other metrics with $6$ isometries, corresponding to spaces that have a constant coefficient for the solid angle element and they have the corresponding Noether invariants, but their physical significance is not quite that obvious. There are no spherically symmetric metrics with only $5$ isometries as was shown in a complete classification of spacetimes by their isometries \cite{aq4}. There are many other spherically symmetric spacetimes with $4$ KVs and any number with only the $3$ of angular momentum, that define spherical symmetry.

Dispensing with spherical symmetry, the Kerr metric represents a spinning point mass, $m$, with angular momentum per unit mass, $a$,
\begin{eqnarray}
& & ds^2 = A dt^2- \frac{\rho^2}{\Delta} dr^2 - \rho^2 d\theta^2 - B^2 d\phi^2 - C dt d\phi~ \, , \nonumber\\ & & \rho^2 = r^2 + a^2\cos^2\theta ~, \quad \Delta = r^2 - 2mr + a^2~, \quad ~A = 1- \frac{2mr}{\rho^2}~ \, ,  \\ & & B = (r^2+a^2) \sin^2 \theta + \frac{2 m r}{\rho^2} a^2 \sin^4 \theta ~, \quad ~C= \frac{2 m r }{\rho^2} a \sin^2\theta \, , \nonumber
\end{eqnarray}
which has two KVs, for time translation and axial rotation. However, it has three Noether symmetries, two of which correspond to energy and angular momentum and an additional one coming from a Killing {\it tensor} (see, for example, \cite{hs}). It has been shown \cite{aq5} that this invariant corresponds to the total angular momentum squared. It is interesting to note that these are the same quantities that are conserved in Quantum Mechanics and that the angular momentum vector is not conserved there either
\cite{aq6}.

\subsubsection{Conformal Symmetries}

\noindent In School Geometry one first learns of congruent triangles and then of similar triangles. The congruent triangles give invariance of the figure under translation, but the similar triangles give it under translation and {\it scaling}, which is provided by changing lengths while leaving angles invariant. This is achieved by scaling the metric tensor, ${\bf g}\rightarrow\tilde{{\bf g}}=\Omega^2({\bf x}){\bf g}$, which is called a {\it conformal transformation}. In this case there will be a {\it conformal Killing vector}, (cKV), or {\it conformal isometry}. Thus, if there is a timelike cKV, though energy conservation is lost, a re-scaled energy is conserved. This applies, for example, to the Friedmann metrics
\begin{equation}
ds^2= dt^2 - a^2(t) \left[ d\chi^2 + f_k^2(\chi) (d\theta^2+\sin^2\theta d\phi^2) \right]~, \label{frw}
\end{equation}
where $k$ gives the normalized constant curvature, being $+1$ for a sphere, $0$ for a plane and $-1$ for a hyperbola; the corresponding $f_k(\chi)$ are $\sin\chi$, $\chi$ and $\sinh\chi$. For the first case the range of $\chi$ is $2\pi$, for the other two it is semi-infinite. In this case there are only
6 isometries as the timelike one is lost, and so energy is not conserved. However, there is a cKV, $a^{-1}(t)\delta^{\mu}_0$, and so energy is conserved up to scaling. The scaling comes from the expansion factor for the Universe, $a(t)$, and the re-scaled conserved quantity is the number of
particles in the expanding volume. In terms of the Lagrangian for the metric tensor, the Lagrangian is not conserved but it is scaled. This yields a {\it conformal Noether invariant}, which is the energy.

\subsubsection{Symmetries of the Electromagnetic Field}

\noindent In the Lagrangian for the electromagnetic field, given immediately after Eq. \eqref{eq20}, the first term gives the source and the second represents the pure electromagnetic field. The corresponding stress-energy tensor for the pure electromagnetic field, given by Eq. \eqref{eq32} is
\begin{equation}
T^{\mu}_{\nu}=F^{\mu\alpha}F_{\nu\alpha}-\frac{1}{4}\delta^{\mu}_{\nu}F^{\alpha \beta}F_{\alpha\beta}~.
\end{equation}
Writing this in terms of the electric and magnetic fields, the Lagrangian is $(E^2-H^2)/8\pi$, for the pure time component of the stress-energy tensor, i.e. the Hamiltonian or energy, we get $(E^2+H^2)/8\pi$, the space-time part is the momentum vector, which gives the {\it Poynting vector}, ${\bf E\times H}/4 \pi$, and the spatial part gives the Maxwell stress tensor
\begin{equation}
\sigma^i_j=E^i E_j + B^i B_j - \frac{1}{2}(E^2+B^2) \delta^i_j~,
\end{equation}
which signifies the symmetric stresses in a shear-free ``fluid''.

Now the 4-gradient of a scalar function can be added to the 4-vector potential, ${\bf A}$ without changing the Maxwell tensor, ${\bf F}$, since the latter is the 4-curl of the former, and the curl-grad vanishes in all dimensions, i.e. if ${\bf A}\rightarrow{\bf\tilde{A}}={\bf A}+{\bf d}f$ then
${\bf F}\rightarrow{\bf\tilde{F}}$. This non-uniqueness of the field is called ``gauge-freedom'' and we say that {\bf A} is invariant under gauge transformations. This freedom must have a group associated with it, and consequently a ``Noether charge''. A conservation associated with it comes from the observation that the second divergence of the Maxwell tensor must be zero, as the second derivative is symmetric but the Maxwell tensor is skew. Hence, the divergence of the first of the Maxwell equations, \eqref{maxwell-eqs}, implies that $j^{\mu}_{\,\, ;\mu}=0$, which amounts to $(\rho v^{\mu})_{;\mu}=0$, which says that the total time derivative of the charge density is zero, i.e. the electric charge is conserved.

For the transformation ${\bf\tilde{A}}=e^{\iota f({\bf x})}{\bf A}$, $\nabla{\bf\tilde{A}}=e^{\iota f({\bf x})}[\nabla{\bf A} + \iota(\nabla f){\bf A}]$. If we now define the so-called ``covariant derivative'', $\tilde{\nabla}=\nabla-\iota\nabla$ then the covariant curl gives ${\bf\tilde{F}}={\bf F}$, so that we recover gauge invariance by multiplying by a position-dependent phase. The multiplication by a phase can be understood geometrically by considering $e^{\iota\, \theta}$ acting on a position vector in the complex plane, i.e. the number $z= x + \iota y$. This transforms to the new point $(x\cos\theta-y\sin\theta) + \iota (y\cos\theta+x\sin\theta)$, which is just a rotation through the angle $\theta$. If $\theta$ is a constant we say that it is a {\it global} gauge transformation and if it is variable a {\it local} gauge transformation. Thus, for the electromagnetic field we have local gauge invariance. The physically relevant symmetry is the local gauge symmetry which yields the conservation of electric charge. Written as the addition of the gradient of a scalar function does not indicate what the associated group is, but in the complex form employed above, this becomes the unitary group in one dimension, $U(1)$. In other words, we can regard Maxwell's theory as the consequence of a $U(1)$ local gauge symmetry, called $U_{em}(1)$. Local gauge theories really came into their own in Quantum Theory (QT), which we shall be seeing in the next section.

\section{Symmetries in Quantum Theory}

\noindent In Quantum Theory (QT) discrete symmetries play a significant role. Since the Lagrangian is left invariant under their action, they are Noether symmetries and will give some conserved quantity. Since they are discrete, they will not reduce the number of variables or order of the equations, but will limit the range. Thus a reflection symmetry halves the range and a global rotational symmetry limits the variable to a semi-closed interval that can be chosen to be $[0,2\pi)$. If there is reflection symmetry, the number of solutions is more than halved (as they must all be even functions; or if there is antisymmetry they must all be odd). This break-up applies to translational shifts, and hence to momentum. However, for angular momentum, we get ${\bf r}\rightarrow-{\bf r}$ and ${\bf p}\rightarrow-{\bf p}$, so ${\bf r}\times{\bf p}\rightarrow{\bf r}\times{\bf p}$. This symmetry is called {\it parity}. Thus, while in a mirror left is converted to right the angular momentum arrow remains unchanged. If a quantity remains unchanged, it is said to have positive parity and if it reverses direction, it is said to have negative parity. It was believed that all fundamental particles have either one or the other (with no cases of neither) and the total parity in any interaction is conserved.

Salam proposed that parity is violated in weak nuclear interactions (due to which there is radioactive decay), and he sent his draft paper for comments to Wolfgang Pauli, a leader of QT and especially of the spin quantum number, which would be reversed if parity is violated. Pauli sent him back the
message, ``Tell my young friend Salam to think of something better.'' He added, ``Parity is conserved --- I can feel it in my bones,'' punning on the use of the phrase ``feel in my bones'', as the phosphorous in the bones would decay if parity is not conserved. The fact is that it {\it is} violated and the phosphorous in the bones {\it does} decay, but the rate of decay is orders of magnitude less than the biological degradation of the bones and could not be noted. Pauli's reasoning was specious and spurious. When Yang and Mills received the Nobel Prize for independently discovering the same
principle, Pauli apologised to Salam, but that did not restore Salam's claim to priority.

There is also time reversal symmetry or asymmetry in any process, in that a film of the process run in reverse would be indistinguishable from the original, e.g. the collision of two billiard balls on a billiards table, or the swing of an ideal pendulum. However, if the surface of the table is
rough, or the pendulum is damped, one can distinguish the forward direction from the reverse. Though non-frictional dynamics obeys time reversal invariance, for fundamental particles it was taken for granted that it is conserved, with positive or negative values. Another property of fundamental
particles is electric charge. It is found that corresponding to each fundamental particle, there is an otherwise identical particle with opposite charge, called an {\it antiparticle}, even though one may be far more common than the other (like electrons and positrons or like protons and antiprotons). Writing the parity reversal operator as P, time-reversal as T and charge conjugation as C, though each one may be separately violated it has been proved that their product, CPT, is conserved in all fundamental processes, (the {\it CPT Theorem}).

More relevant for recent developments, are continuous symmetries. Recall that QT uses a complex ``wave function'', satisfying the Schr\"odinger, Klein-Gordon, or Dirac equations. It gives a complex amplitude whose magnitude square represents the probability of finding a quantum entity at
some place at some time. (In fact, even for the classical electromagnetic field theory, it is convenient to use complex variables, see e.g. \cite{ll,jdj}.) When the classical electromagnetic field is ``quantized'' by Dirac's procedure (see e.g. \cite{jjs}), the 4-vector field corresponds to a
spin-one particle, called the {\it photon}. If the gravitational field could be quantized, it would correspond to a spin-two field that people call the {\it graviton}. More generally, a field represented by a tensor of rank $n$ corresponds to spin-$n$ quantized field a. These have a real representation but, Dirac showed that there would be half-integer spin fields as well, such
as the electron, represented by a 4-dimensional complex vector of the representation space of $U_{em}(1)$, called a {\it Dirac spinor}. Spin-$n/2$ has an $n$-index spinor. The usual vector corresponds to a two index spinor, or a single index tensor. Note that $U(1)$ is locally isomorphic to $SO(2)$, as they both give rotations, but the former is a double covering of the latter because of the complex representation.

While rotations in two dimensions commute, and so $SO(2)$ and $U(1)$ are Abelian, they do not in general and so are non-Abelian. As $U(n)=SU(n)\bigotimes U(1)$, the next simplest unitary group is $SU(2)$. Like the transpose of an orthogonal matrix is its inverse, the Hermitian conjugate
of a unitary matrix is its inverse. Since a $2\times2$ matrix has four complex entries subject to four real constraints, $U(2)$ has four independent parameters and hence $SU(2)$ has three. As 3-d rotations are also three, $SU(2)$ is locally isomorphic to $SO(3)$, again with a double covering. This
is the symmetry group for weak interactions at high enough energy $\sim150~GeV$, as shown by Glashow, Salam and Weinberg \cite{as}. In fact, the weak and electromagnetic forces are unified at energies $\sim150 GeV$ but the {\it electroweak} (EW) symmetry group, $SU_W(2)\bigotimes U_Y(1)$,
breaks down at lower energies to the usual $U_{em}(1)$ below that. The ``$Y$'' is the conserved {\it hypercharge}, which mixes the charge of electromagnetism with that of the weak nuclear force, on account of which the photon of electromagnetism is a mixture of the bare $U_{em}(1)$ and the
neutral component of the $SU_Y(2)$, called a {\it neutral current} (the other two being charged currents), above the unification energy. Murray Gell-Mann \cite{mgm} had proposed an $SU(3)$ gauge group for three {\it quarks}, for the strong nuclear force. Originally an energy-dependent term not respecting the symmetry, was inserted in the Lagrangian to break the symmetry, which is
negligible at lower energies, but dominates at higher energies, and so the symmetry breaks --- gradually. Peter Higgs \cite{pwh} suggested a mechanism whereby the symmetry breaks spontaneously at a critical energy, due to a field with a vacuum expectation value at higher energies that acquires a mass and becomes a physical spin zero particle, called a ``Higgs boson'', as the energy drops below its mass. This mechanism was used by Salam to develop the ``electroweak'' unified theory, and was then used for $SU(3)$-breaking. The new conserved quantum number in this case was called ``colour'', and the
resulting theory called {\it quantum chromo-dynamics} (QCD), denoted by $SU_c(3)$. It has eight generators and that yields eight gauge bosons called ``gluons''. The {\it standard model} of particle physics is, then, $SU_C(3)\bigotimes SU_W(2)\bigotimes U_Y(1)$.

\subsection{Gauge Grand Unification Symmetry}

\noindent The critical energy of the standard model is the same as the EW theory, but it consists of three forces with very different strengths: (a) weak $\sim10^{-8}$; (b) electromagnetism $\sim10^{-2}$; and (c) strong $\sim1$. As the interaction energy increases the strengths change; the weak and electromagnetic becoming stronger at relative rates that go inversely as the relative strengths and the strong gets weaker. At first it appeared that all three should meet at somewhere around $10^{12}-10^{15} GeV$. Since the Universe is cooling as it expands (like the gas in a refrigerator), this was taken to indicate that there may have been a time in the early stages of the Universe when all three were unified and this ``grand unified theory'' (GUT) broke at the critical energy to yield the Universe as we see it now. In that case we would need a bigger group that would break down to the groups of the
standard model.

In 1971 Jogesh Pati and Abdus Salam proposed an $SU_C(4)\bigotimes SU_f(4)$ GUT (published in 1974 \cite{ps}), with the weakly interacting particles (called leptons) as a fourth colour prior to symmetry breaking, and with four ``flavours'' of the leptons/quarks, $(u,d,e,\nu_e)$ giving the the usual
protons, neutrons and electrons and a more massive set $(s,c,\mu,\nu_{\mu})$, that was known at the time. Whereas Gell-Mann used fractional charges for his quarks: $+2/3$ for $u$; $-1/3$ for $d$; $-3/3$ for $e$; and $0/3$ for $\nu_e$, Pati and Salam used integer charges. but it has an enormous $225$
generators! The smallest simple group containing the full standard model is $SU(5)$, which could break into the standard model at some critical energy, taken to be $\sim10^{15}GeV$ and there would be a $\sim10^{15}GeV$ Higgs boson associated with that unification. This was proposed in 1974 by Howard
Georgi and Sheldon Glashow \cite{gg}. The number of generators would be 24 in this case. By this time there was reason to believe that there were three sets of the basic four particles, and the two heavier sets were not needed for the unified theory, but came as redundant copies. The one set had three
colours for the quarks and only one each for the leptons. However, the theory took left-handed and right-handed spins for the quarks and the electron, but only a left-handed version of the neutrino, $\nu_e$. This unaesthetic break-up was put in a cumbersome way into two representations of the group.
To avoid this problem in 1975 Harald Fritzsch and Peter Minkowski \cite{fm} proposed an $SO(10)$ theory that put all the particles into a single multiplet, but at the expense of expanding the number of generators to 45. While the $SU(5)$ breaks in a single step, $SO(10)$ can break into a
left-right symmetric model, which then breaks down into the standard model.

More precise experiments showed that the three strengths do not come together at a single energy, with the extrapolations forming a triangle. As such, the basic raison d'etre was lost and the whole unification enterprise seemed to be in serious jeopardy. Further, since the quarks and leptons could
inter-convert at sufficiently high energies, the proton would be unstable. The predictions for proton decay of $SU_C(4)\bigotimes SU_f(4)$ and $SU(5)$ were experimentally violated and the only reason $SO(10)$ escaped was that it did not have a definite prediction --- which is hardly a recommendation.
Something more was needed to save GUTs.

\subsection{Supersymmetry and Unification}

\noindent Quantum Field Theory (QFT) had problems since its inception. Taken beyond the lowest level, the calculations for any interaction yield infinite probabilities, called ``divergences''. Since the wave functions are unit norm vectors in a Hilbert space, it is argued that if they seem to become infinite
they should be ``renormalized''. However, it turns out that all theories, especially gravity, cannot be renormalized. Renormalizing infinity seems suspect to many in any case. A finite theory, which can then be renormalized in a meaningful way is needed. A method came from a novel proposal of treating spinors and tensors as different representations of a unified symmetry, but using commutators for the products of tensors and {\it anti-commutators} for the products of spinors (or multi-spinors), which complicates the Lie algebra to what is called a ``super-algebra''. This {\it supersymmetry} (SUSY) was proposed by Julius Wess and Bruno Zumino \cite{wz}. Their cumbersome formalism was put into a more usable form by Abdus Salam and his student, John Strathdee \cite{ss}.

Tensors correspond to fields with integer spins in units of Planck's constant (divided by $2\pi$), $\hbar$, while spinors correspond to half-integer spins. The thermodynamic distributions, giving the speeds or energies of the gas particles for the former are called {\it bosons} and for the latter {\it fermions}, are different at low temperatures but behave in much the same way at higher temperatures. In no way could {\it this} be taken to be the unification talked of. However, there is no spontaneous symmetry breaking mechanism for it either. Nevertheless, assuming that the symmetry {\it does} apply in the sense of a fundamental theory at some higher energy and then breaks down, it would modify the standard model significantly by introducing an extra parameter of the symmetry-breaking energy. In the standard model the number of Higgs bosons is not constrained. For definiteness one takes the minimal number of Higgs fields, and this is called the {\it minimal standard model} (MSM). With SUSY one gets an MSSM. It was found that the three constants came together at $10^{16}GeV$ if SUSY is assumed to be broken at $1TeV=10^3Gev$. Thus, if SUSY is to save GUTs, it must be seen to break at this energy, which was reached long ago at the Large Hadron Collider at CERN. It has not been seen so far, and people are jumping through hoops to keep SUSY alive, but she is on life-support.

Could a higher unification of {\it all} forces save the day? The divergences of the bosons seemed to be canceled at the lowest non-trivial level of calculation of interaction cross-sections, by those of its fermionic super-partner, and vice versa. Gravity is a non-renormalizable theory. With SUSY it is called {\it supergravity} (SUGRA). The super-partner of the hypothetical ``graviton'' was a spin 3/2 field called a {\it gravitino}. In SUGRA it turned out that at the next non-trivial level the divergences {\it did not} cancel. This problem was resolved by using an {\it extended} SUGRA,
in which a second gravitino was inserted. This mechanism worked by introducing a new gravitino at the next level up to the eighth level, because of the extra parameter inserted, but by the same token it will not work beyond that \cite{pvn}. One also needed to go to higher dimensions of spacetime and strings or membranes instead of point particles \cite{sssm}.

\subsection{Twistor Quantization and Unification}

\noindent A totally different approach to combine GR and QT was proposed by Roger Penrose. As he saw it, the problem with combining the two is that since the field to be quantized is the metric tensor, which defines the distance between two {\it points}, once the quantization is done the spacetime points
will no longer form a continuum, but will be discrete. Thus, we would no longer be able to use Calculus for Geometry. He pointed out that the quantity that we know is quantized, and we know how to deal with, is angular momentum or spin angular momentum. Regarding particles as just a collection of spins,
one could only {\it know} of the existence of the other by exchanging a unit of spin, $\hbar/2$. He wrote a paper entitled Spin Networks in 1967 or 1968, which he gave to one of us (AQ), when he was suggesting possible lines for AQ's PhD research (when he gave one of us, AQ, an unpublished paper on {\it spin networks} to see if he wanted to work on the idea for his PhD). It was later published in 1971
\cite{rp1}. He regarded the bundles of spin as moving at the speed of light, so that he could express them in terms of {\it spinors} (for which he had given a geometrical visualization \cite{rp2}).

Spinors can be thought of as vectors of the representation space of the {\it symplectic group}, $Sp(2n)$, over $\mathbb{R}$. This group leaves invariant the null structure, so that all vectors have zero magnitude. Penrose was dealing with {\it two-component} spinors, so that the symplectic structure is provided by the Levi-Civita symbol, $\epsilon_{AB}~(A,B=0,1)$, which is 0 when $A=B$, 1
when $A=0,B=1$ and $-1$ when $A=1,B=0$. Using the set of four Pauli spin matrices $\sigma_{\mu}^{AA'}$, a vector, $x^{\mu}$ can then be written as a complex matrix, $x^{AA'}=x^{\mu}\sigma_{\mu}^{AA'}$. If the vector is null then we can write $x^{AA'}=\xi^A\overline{\xi}^{A'}$. Here $\xi^A$ is a two-component spinor (for spin vector) and can be visualised as a flagpole lying along the null cone with a half-plane element stuck on it like a pennant. The symmetry group for it is $SU(2)$. Multiplying the spinor by a
complex number scales the flagpole by the magnitude of the factor and the pennant is rotated through twice the argument of the factor. This explains why $SU(2)$ is a double covering of $SO(3)$, since the scaling by $e^{\iota\pi}$ will leave the vector unchanged but will reverse the direction
of the spinor.

The spinor representation of the covariant derivative operator $\nabla_{\mu}$ is $\nabla_{AA'}$ , and the Killing equation can be written for a spinor as $\nabla^{AC'}\xi^B+\nabla^{BC'}\xi^A=0$, which is called the {\it twistor equation} \cite{rp3}. The twistor contains the information of the spinor,
$\xi^A$ and an $\eta^{A'}=\iota x^{AA'}\xi_A$. Thus the twistor is given by the pair of spinors, $Z^{\alpha}=(\xi^A,\eta^{A'}),~(\alpha =0,...,3)$. Single or multi-index twistors are solutions of the zero rest-mass field equations \cite{rp4}, with the number of indices corresponding to the spin of
the field. A twistor has four complex components, corresponding to eight real components, but the relevant information does not depend on the overall magnitude, as the position can go on sliding up the flagpole out to infinity. Thus the twistor is an entire null ray. As such one only needs the projective
space of twistors, which is three complex, or six real, dimensional. The $x^{\mu}$ in the twistor will generally be complex and the {\it twist} in the congruence of geodesics (given by the relevant spin coefficient, see \cite{pr}) is either positive or negative. The corresponding twistors are
said to belong to $C^+$ or $C^-$. However, when $x^{\mu}$ is real, there is one extra constraint and there are only five real components left. Such twistors are called {\it null twistors}. Hence there is a five-dimensional hypersurface, $N$, separating the positive and negative projective twistor
spaces, $C^+$ and $C^-$. Elements of $N$ correspond to entire real null rays in real Minkowski space, and the congruence of such rays is also five dimensional. The Penrose transform between the twistor space and complexified Minkowski space signifies a duality between the two.

We can generate solutions of the zero rest-mass field equations by using certain ``intertwining integrals'' \cite{aq7,aq8}. This enables us to do contour integration over $N$, for twistor fields, which will start in one of the six-dimensional spaces, pass through $N$ and then go back into the original one. This procedure yields scattering amplitudes, and hence probabilities, for the various scattering processes in the high energy limit \cite{rp5,aq9}. Whereas standard QFT yields infinite probabilities, and SUSY, SUGRA, superstrings and supermembranes hope to achieve cancellations of the infinities, the twistor approach yields finite answers automatically, and the renormalization involved is only division by a magnitude obtained by summing a convergent series of finite magnitude terms.

\section{Complex Lie and Noether Symmetries}

In his work, Lie had considered the transformation of the independent and dependent variable by differentiable transformations, called {\it point transformations}, as the space of the variables can be thought of as two-dimensional, one for the independent and one for the dependent, with the
specific values of the variables represented by a point. When the variables are transformed the points get shifted. Thus $(x,y)\rightarrow(\tilde{x}(x,y),\tilde{y}(x,y))$ is a point transformation.
It is particularly useful to make the differentiability explicit, by using the transformations in infinitesimal form, so that one sees a smooth path traced out by the moving point. Thus we can write
\begin{equation}
\tilde{x}= x + \epsilon\xi(x,y) + O(\epsilon^2)~,~\tilde{y} = y + \epsilon \eta(x,y) + O(\epsilon^2)~,
\end{equation}
where $\epsilon>0$ is an infinitesimal. The {\it infinitesimal generator} of the transformation is then
\begin{equation}
{\bf X}=\xi(x,y)\frac{\partial}{\partial x} + \eta(x,y))\frac{\partial}{\partial y}~.
\end{equation}
A scalar ODE is said to be {\it symmetric} or {\it invariant} under a transformation if the graph of its solution is preserved by the transformation. He provided various methods to check whether a given system of differential equations is invariant under a given point transformation and others were developed later, see e.g. \cite{hs}. In particular, he discussed when the equations could be converted to linear form by point transformations, called {\it linearization}, and thereby solved comparatively
easily. He especially studied the criteria for scalar second order ODEs to be so transformed \cite{sl3}, see e.g \cite{bk}. Despite the importance of PDEs in applications (especially in Fluid Dynamics), we restrict our attention to ODES. The reason is that while the general solution of an ODE is unique up to (at most) as many arbitrary constants as the order of the equation, PDEs,
have infinitely many solutions. The definite statements available for ODEs are lost for PDEs.

For second order ODEs, $y^{\prime\prime}= w(x,y,y^{\prime})$ we need to include the first derivative as if it were an independent variable and use a 3-d space. For higher order scalar ODEs the space has to be {\it extended} (or {\it prolonged}) to include all derivatives up to the next to highest
derivative, as if all are independent of each other. The graph of the solution must then remain invariant in the projected two-dimensional space. Thus for the second order scalar ODE the prolonged infinitesimal generator is
\begin{equation}
{\bf X}^{[1]}=\xi(x,y)\frac{\partial}{\partial x}+\eta(x,y))\frac{\partial}
{\partial y}+\eta^{[1]}(x,y,y^{\prime}))\frac{\partial}{\partial y^{\prime}}
~,
\end{equation}
where $\eta^{[1]}$ is given in terms of $\eta$ and the derivative of $\xi$. As the  order increases the generator gets prolonged further by adding another term involving the derivative with respect to the next order derivative. For an $n^{th}$ order ODE we need to prolong up to $\eta^{[n-1]}$, where
\begin{equation}
\eta^{[n-1]}(x,y, \ldots, y^{(n-1)})=
\frac{d}{dx}\eta^{[n-2]}(x,y,\ldots, y^{(n-3)})-y^{(n-1)}\frac{d}{dx}\xi(x,y)~.
\end{equation}

Lie had proved that second order scalar ODEs are linearizable only if they are at most cubically semilinear and the four coefficients satisfy a set of four first derivative constraints involving two arbitrary functions. Tresse \cite{at} reduced them to two second order constraints without the arbitrary functions. Note that these equations do not have to be {\it solved} but only {\it checked}. Methods were later developed to simply write down the solution of second order quadratically semilinear linearizable systems \cite{mq1} and were later generalized to the cubically semilinear case \cite{mq2,mq22}.

Lie used complex functions of complex variables for his analysis of differential equations. Of course, he had to take the functions to be not only continuous but differentiable. Now complex differentiability implies analyticity. He did not explicitly use this fact in his analysis. However, it obviously has significant consequences, as the dependent and independent variables will be constrained by the Cauchy-Riemann (CR) equations. Thus, even a scalar first order ODE, split into its real and imaginary parts, is actually a system of four first order PDEs. This point was noted and
exploited by Ali, Mahomed and Qadir \cite{amq1,amq12}, who called it {\it complex symmetry analysis} (CSA). If the dependent variable is $w=u+\iota v$ and independent variable $z=x+\iota y$, one function of one variable yields four functions of two variables, and so a system of PDEs. For the resulting system to be of ODEs, one must restrict the independent variable to the real part, $x$, only. Thus the complex scalar ODE will now correspond to a system of two ODEs along with a set of CR equations. A first order ODE will then split into the two first order ODEs, $u^{\prime}=f(x,u,v)$ and $v^{\prime}=g(x,u,v)$, by
taking the real and imaginary parts of the equation. The CR equations for the corresponding system will be $f_u=g_v~,~f_v=-g_u$. It is this system that would now be analysed. The formalism is easily extended to higher order ODEs.

The complex generator ${\bf W}$ splits into its real and imaginary parts, as
\begin{eqnarray}
{\bf W}&=&\xi(x,w)\frac{\partial}{\partial x}+
\eta(x,w))\frac{\partial}{\partial w} \nonumber \\
&=&\xi^r(x,u,v)\frac{\partial}{\partial x} \nonumber \\
&+&\frac{1}{2}[(\eta^r(x,u,v)\frac{\partial}{\partial u}+
\eta^i(x,u,v)\frac{\partial}{\partial v})+
\iota(\eta^i(x,u,v)\frac{\partial}{\partial u}-
\eta^r(x,u,v)\frac{\partial}{\partial v})] \nonumber \\
&=&{\bf X}+\iota{\bf Y}~,
\end{eqnarray}
the half coming from the requirement that ${\bf W}w=1$. The prolongation is, of course, still more cumbersome to write, but easy to obtain. Ali, Mahomed and Qadir applied their CSA to the linearization of second order scalar ODEs \cite{amq2}, and found that they could linearize the complex scalar second
order ODE corresponding to a 2-d non-linearizable system. In fact, it was found that systems with less than the eight generators required for a single scalar second order ODE may correspond to a complex scalar linearizable ODE. Further, systems without enough infinitesimal symmetry generators to be
solvable by symmetry methods, there was an example of a second order 2-d system that has {\it no} generator, but corresponds to a complex scalar second order ODE, and its solution was obtained \cite{sqa,sqa1,sqa2}.

Extending CSA to Lagrangians to be able to use it for Noether symmetries is non-trivial. The reason is that the Lagrangian is necessarily defined for the real domain. The extension presents us with two basic problems that may be thought of as two faces of the same coin. The Lagrangian is the kernel of a
functional, and functionals map functions into $\mathbb{R}$ not $\mathbb{C}$. The whole purpose of defining it is to find the form of the dependent variable for which the functional takes a minimum value. As such the image space for it has to be an ordered set like $\mathbb{R}$, and not a partially ordered set like $\mathbb{C}$. The first problem is simply dealt with by fiat, redefining the range of the functional to be $\mathbb{C}$. The associated problem is dealt with by using the {\it magnitude} of the functional to be minimal, rather the functional itself \cite{amq3}.

It turned out that invariants could be obtained for the complex Lagrangians and provide new insights into the physical significance of the Noether invariant \cite{faq}. In particular, when a complex scalar harmonic oscillator equation is split, it gives a system of coupled harmonic oscillators and the Noether invariant gives the energy in each oscillator {\it and identifiably separately in the field between them}. This was expressed as ``seeing the energy in the field through complex glasses''. It
is easy to extend to the time dependent harmonic oscillator and see how the energy in the coupled oscillators and in the field are transferred about. It turns out that the complex Noether symmetries do not provide any new invariants, but they {\it do} provide them more easily and put them into
different combinations that may be  more insightful \cite{sqf}.


\section{Concluding Remarks}

Mathematically, Noether symmetries provide double reduction of the order and/or the number of variables in a differential equation. One of the methods of achieving a reduction is by determining an invariant combination of the dependent and independent variables and their derivatives. Writing the
invariant as an arbitrary constant, it can be used to write, say, the highest derivative in the combination, in terms of the  other variables. It is particularly needed for PDEs, as ``solving them'' without the boundary conditions is not very meaningful, and the invariants can incorporate the
boundary conditions. While other methods could be used for the same mathematical purpose, they cannot provide the physical insights of the ``Noether charge'' that the invariants do. In this paper the physical insights obtained from Lie and Noether symmetries and invariants was reviewed. Worth special mention is the use of Noether symmetries in Quantum Theory, which has not received enough attention from those working in Symmetry Analysis. Since one of the most important outstanding problems in
fundamental Physics is the unification of Quantum Theory and General Relativity, the introduction of new methods may lead in solving it. One such method is the explicit use of complex analyticity of Lie groups that was also briefly mentioned, but there are many others.

There is considerable activity in spacetime symmetries \cite{gsh}, including not only isometries (also called KVs), but also the scaling symmetry of the metric tensor, called a {\it homothety} (the special case of a constant conformal factor). There is also interest in the symmetries of the Ricci ({\bf Ric}), Riemann ({\bf Rie}) and Weyl ({\bf C}) tensors, which are called {\it Ricci}, {\it Riemann} and {\it Weyl collineations}, given by $\pounds_{\bf k}{\bf Ric}=0$, $\pounds_{\bf k}{\bf Rie}=0$ and $\pounds_{\bf k}{\bf C}=0$. Much of it is on their physical significance \cite{tm1990,bq1993,hall1996,bqaa,cb2002,bkq,hqs} and many problems continue to arise, which would be worth exploring. A different line was developed of obtaining all metrics with isometry groups containing a minimal group, which was called a ``complete classification'' of spacetimes with the minimal symmetry \cite{bq1987,qz1995,dz1997}. The idea was to be able to pick up spacetimes that have the desired symmetry and use the metric to obtain the stress-energy tensor by constructing the Einstein tensor for it. Thus one manages to ``solve the Einstein equations without having to solve them''. This was initially done for isometries of spherically symmetric, static metrics, but was then extended to isometry groups of only three dimensions. It was further extended to homotheties and collineations. Classification by Noether symmetries \cite{feroze2,feroze3,feroze4,feroze5,feroze6,jamil1,jamil2} has yielded solutions of the Einstein equations along with their conserved quantities. This line is also worth pursuing. A more difficult problem is to completely classify by a two-dimensional isometry group, like cylindrical symmetry. If that is done, it may be possible to extend it to homotheties and
collineations.

As mentioned in Section 7, there is great need for workers in Lie symmetry analysis to enter into QFT, as the work using symmetries is pursued by physicists who are not generally well-versed in the methods developed by Lie and subsequent workers. (However, it is not enough for the workers to simply
take an open problem in QFT, without understanding what is entailed in it. It would be all too easy to get spurious and irrelevant results if the context is not understood.) In particular, any new solutions of the Newman-Penrose equations \cite{rp2,pr} would be {\it extremely useful} as they come equipped
with their physical significance. Since they deal with invariants a Noether symmetry formulation for them may be possible, and would be a major contribution.

Coming to Section 8, the field is wide open. Very powerful complex methods have been developed, where they can be applied. One direction to go in following this up, is to check open problems to see if the powerful complex methods can be used there or not. Unfortunately, one runs the risk of wasting
a lot of time finding that the methods are {\it not} applicable to the chosen problem. One must also bear in mind that while CSA provides solutions of systems of ODEs that cannot be solved by traditional symmetry methods, Noether symmetries will not give new invariants, but only new combinations of
the old invariants. As pointed out above, they {\it can} provide new insights into the physical significance of the invariants. A very much more important line of work to follow in this area, is to find an explanation of {\it why} the complex methods provide answers, where they do. If one can answer that question, one should be able to formulate criteria for the applicability of complex methods and thus avoid wasting time on searching for problems where complex methods can be applied and go directly to those where it can; or perhaps even find ways to tweak the methods to make them more generally
applicable.

\section*{Acknowledgements}

We would like to thank unknown referees for their valuable comments and suggestions.



\end{document}